\documentclass[
superscriptaddress,aip,apl,reprint,twocolumn,amsmath,amssymb,showpacs]{revtex4-2}
\usepackage{}

\usepackage[dvipsnames]{xcolor} 

\usepackage{graphicx}  
\usepackage{mathptmx} 
\usepackage{epstopdf}
\usepackage{dcolumn}   
\usepackage{bm}        
\usepackage{amssymb}   
\usepackage{amsmath}   
\usepackage{amsthm}
\usepackage{chemarrow}
\usepackage{color}
\usepackage{mathrsfs}
\usepackage{float}
\usepackage{bbold}
\usepackage{bbm}
\usepackage[a4paper,colorlinks=true,
linkcolor=blue,citecolor=blue,
pdfauthor={ },
pdftitle={ },
pdfsubject={ },
pdfkeywords={ }]{hyperref}

\theoremstyle{plain}
\newtheorem*{theorem*}{Theorem}

\begin{document}


\title{Cooperative Spin Amplification}

\date{\today}

\author{Minxiang Xu}
\affiliation{
CAS Key Laboratory of Microscale Magnetic Resonance and School of Physical Sciences, University of Science and Technology of China, Hefei, Anhui 230026, China}
\affiliation{
CAS Center for Excellence in Quantum Information and Quantum Physics, University of Science and Technology of China, Hefei, Anhui 230026, China}
\affiliation{
\mbox{Hefei National Laboratory, University of Science and Technology of China, Hefei 230088, China}}

\author{Min Jiang}
\email[]{dxjm@ustc.edu.cn}
\affiliation{
CAS Key Laboratory of Microscale Magnetic Resonance and School of Physical Sciences, University of Science and Technology of China, Hefei, Anhui 230026, China}
\affiliation{
CAS Center for Excellence in Quantum Information and Quantum Physics, University of Science and Technology of China, Hefei, Anhui 230026, China}
\affiliation{
\mbox{Hefei National Laboratory, University of Science and Technology of China, Hefei 230088, China}}

\author{Yuanhong Wang}
\affiliation{
CAS Key Laboratory of Microscale Magnetic Resonance and School of Physical Sciences, University of Science and Technology of China, Hefei, Anhui 230026, China}
\affiliation{
CAS Center for Excellence in Quantum Information and Quantum Physics, University of Science and Technology of China, Hefei, Anhui 230026, China}
\affiliation{
\mbox{Hefei National Laboratory, University of Science and Technology of China, Hefei 230088, China}}

\author{Haowen Su}
\affiliation{
CAS Key Laboratory of Microscale Magnetic Resonance and School of Physical Sciences, University of Science and Technology of China, Hefei, Anhui 230026, China}
\affiliation{
CAS Center for Excellence in Quantum Information and Quantum Physics, University of Science and Technology of China, Hefei, Anhui 230026, China}
\affiliation{
\mbox{Hefei National Laboratory, University of Science and Technology of China, Hefei 230088, China}}

\author{Ying Huang}
\affiliation{
CAS Key Laboratory of Microscale Magnetic Resonance and School of Physical Sciences, University of Science and Technology of China, Hefei, Anhui 230026, China}
\affiliation{
CAS Center for Excellence in Quantum Information and Quantum Physics, University of Science and Technology of China, Hefei, Anhui 230026, China}
\affiliation{
\mbox{Hefei National Laboratory, University of Science and Technology of China, Hefei 230088, China}}

\author{\mbox{Xinhua Peng}}
\email[]{xhpeng@ustc.edu.cn}
\affiliation{
CAS Key Laboratory of Microscale Magnetic Resonance and School of Physical Sciences, University of Science and Technology of China, Hefei, Anhui 230026, China}
\affiliation{
CAS Center for Excellence in Quantum Information and Quantum Physics, University of Science and Technology of China, Hefei, Anhui 230026, China}
\affiliation{
\mbox{Hefei National Laboratory, University of Science and Technology of China, Hefei 230088, China}}

\begin{abstract}
Quantum amplification is recognized as a key resource for precision measurements.
However, most conventional paradigms employ an ensemble of independent particles that usually limit the performance of quantum amplification in gain, spectral linewidth, etc.
Here we demonstrate a new signal amplification using cooperative $^{129}$Xe nuclear spins embedded within a feedback circuit, where the noble-gas spin coherence time is enhanced by at least one order of magnitude.
Using such a technique, magnetic field can be substantially pre-enhanced by more than three orders and is in situ readout with an embedded $^{87}$Rb magnetometer.
We realize an ultrahigh magnetic sensitivity of 4.0\,fT/Hz$^{1/2}$ that surpasses the photon-shot noise and even below the spin-projection noise of the embedded atomic magnetometer, allowing for exciting applications including searches for dark matter with sensitivity well beyond supernova constraints.
Our findings extend the physics of quantum amplification to cooperative spin systems and can be generalized to a wide variety of existing sensors, enabling a new class of “cooperative quantum sensors”.
\end{abstract}

\maketitle

Quantum amplification that offers the capability of enhancing weak signals is ubiquitous and essential to various frontiers of science \cite{clerk2010introduction},
ranging from ultrasensitive magnetic and electric field sensing \cite{macklin2015near,kotler2011single, boss2017quantum},
mechanical oscillator motion measurements \cite{burd2019quantum},
and optical amplifiers \cite{akiyama2007quantum, zavatta2011high} to determination of fundamental constants \cite{rosi2014precision}, frequency standards \cite{diddams2004standards}, and searches for dark matter \cite{bradley2003microwave, budker2014proposal, Jiang2021} and exotic forces beyond the standard model \cite{su2021search}.
To date,
the well-established paradigm of quantum amplification is mostly based on using independent quantum systems,
including superconducting qubits \cite{macklin2015near}, atomic and molecular spins \cite{budker2014proposal, Jiang2021, su2021search}, photons \cite{akiyama2007quantum, zavatta2011high}, nitrogen-vacancy centers in diamonds \cite{boss2017quantum, breeze2018continuous}, trapped-ion qubits \cite{kotler2011single, burd2021quantum}, etc.
The individuals in independent systems amplify the measured signal independently and the total response is the summation of individuals, which in practice leads to limits on the performance of quantum amplifiers,
including operation frequency, spectral linewidth, and gain.

Cooperative systems have recently attracted extensive attention and provided opportunities for novel applications \cite{suefke2017hydrogen, reimann2015cavity, scheibner2007superradiance, jiang2021floquet, shahmoon2017cooperative, rui2020subradiant, norcia2018, wang2019investigating, jin2015proposal}.
In contrast to independent systems, the individuals in cooperative systems experience each other and their evolution depends on the state of the entirety.
Various experimental systems have explored the rich phenomena of cooperative systems, for example, cooperative emitting \cite{oxborrow2012room, suefke2017hydrogen, reimann2015cavity, scheibner2007superradiance, jiang2021floquet} and scattering \cite{shahmoon2017cooperative, rui2020subradiant}, one-axis-twisting dynamics \cite{norcia2018}, and spectral narrowing \cite{wang2019investigating, tang2023pt}.
Cooperative systems could be a promising platform to explore advanced quantum amplification beyond independent systems, partially because such systems provide an ideal way to engineer the coherence time of quantum systems and thus enhance signal response.
The combination of cooperative systems and quantum amplification may open up exciting opportunities for developing new quantum amplifiers with improved performance, especially in gain.
Such amplifiers would find promising applications in precision measurements, for example, ultrasensitive magnetometers \cite{trifunovic2015high, su2022review}, magnetoencephalography \cite{iivanainen2019scalp, boto2016potential}, geomagnetic anomaly detection \cite{sheinker2009magnetic}, and searches for new physics beyond the standard model \cite{Jiang2021, su2021search}.

In this Article, we demonstrate a new magnetic-field signal amplification using cooperative noble-gas nuclear spins.
In experiment, we prepare cooperative $^{129}$Xe spins by acquiring the $^{129}$Xe signal with an embedded $^{87}$Rb magnetometer and then feeding the signal back to the $^{129}$Xe spins with a feedback circuit.
Our investigation shows the dynamics under different feedback strength.
The nuclear-spin coherence time is significantly prolonged by more than one order of magnitude, and 2400-fold improvement in signal amplification is realized using such cooperative spins.
We name these collective phenomena as ``cooperative amplification".
As a first application, our approach constitutes a new technology for enhancing and measuring magnetic fields with a sensitivity of 4.0\,fT/Hz$^{1/2}$,
which surpasses photon-shot noise and even spin-projection noise of the embedded $^{87}$Rb magnetometer.
It is noteworthy that this quantum-enhanced measurement scheme does not rely on entanglement \cite{bao2020}.
We discuss the promising applications of our amplification technique in the searches for hypothetical particles with a sensitivity well beyond the stringent supernova constraints \cite{1978JETPL, Raffelt2008}.
The present amplification technique should be generic for a wide range of sensors and constitute a new class of cooperative sensors.

\begin{figure*}[t]  
	\makeatletter
\centering
	\def\@captype{figure}
	\makeatother
	\includegraphics[scale=1.0]{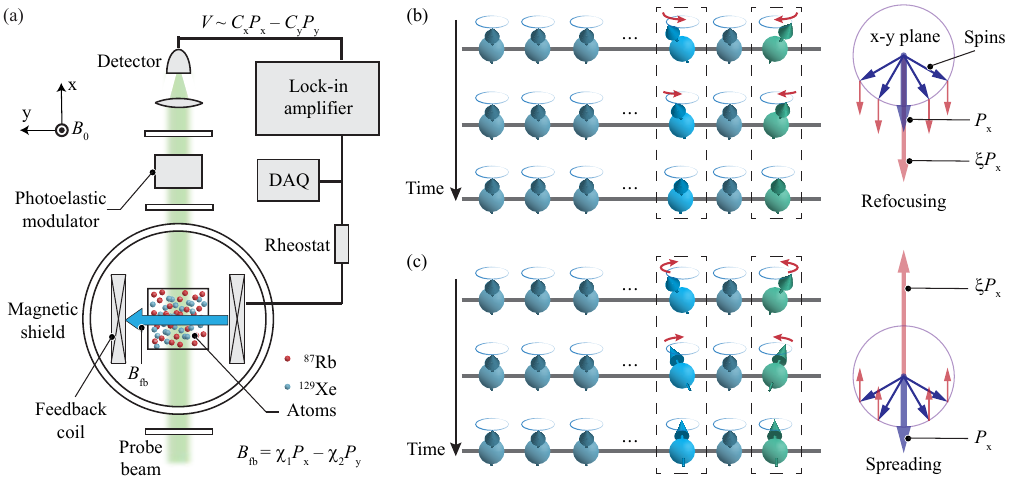}
	\caption{Setup and conceptual description of cooperative dynamics. (a) Sketch of experimental setup. The polarization and probing of $^{129}$Xe atoms are achieved through spin-exchange collisions with $^{87}$Rb atoms. Real-time feedback is provided to the system via a feedback coil. The feedback field includes $P_x$ and $P_y$ signals of $^{129}$Xe. The amplitude of the feedback is controlled by an adjustable rheostat, and the sign is controlled by the connecting polarity. A bias field $B_0$ is applied along the pumping direction. The diagram does not include the pump beam. (b) Refocusing effect in positive feedback mode. Some spins experience dephasing at certain points in time (highlighted with bright colors). The feedback field applies a torque on the dephased spins, causing them to reorient and refocus towards the collective spin. The right inset illustrates the spin dynamics. Each individual spin undergoes a torque (indicated by red arrows) parallel to the collective spin. As a result, the dephased spins tend to refocus, leading to an effective enhancement of the coherence time. Precession is omitted in the dynamical diagram. (c) Spreading effect in negative feedback mode. In this mode, the feedback-induced torque is anti-parallel to the collective spin, causing the dephased spins to align in the opposite direction. Consequently, the effective coherence time decreases as the dephased spins deviate from the collective spin.}
	\label{figure1}
\end{figure*}

Our experiments are carried out in a setup similar to that of Refs.\,[\onlinecite{jiang2021floquet}, \onlinecite{jiang2022floquet}], as depicted in Fig.\,\ref{figure1}(a).
A 0.5\,cm$^3$ cubic vapor cell contains 20\,torr $^{129}$Xe, 250\,torr N$_2$, and a droplet of enriched $^{87}$Rb.
The $^{129}$Xe spins are polarized through spin-exchange collision with optically pumped $^{87}$Rb atoms, as there are no optical transitions available for $^{129}$Xe spins from the ground levels.
A bias field $B_0$ is applied along the pumping direction (the $z$ axis).
The two steps, i.e. measurement and feedback, establish the indirect interaction among spins.
The $^{129}$Xe nuclear magnetization generates an effective magnetic field $\mathbf{B}_\textrm{eff}=\lambda M_0 \mathbf{P}$ on $^{87}$Rb atoms through Fermi-contact collisions \cite{walker1997spin, gentile2017optically}, where $\lambda=8\pi\kappa_0/3$ is the Fermi-enhancement factor, $\kappa_0 \approx 540$ for $^{87}$Rb-$^{129}$Xe system, $M_0$ is the maximum magnetization of the $^{129}$Xe with unity polarization, $\mathbf{P}$ is the equilibrium polarization vector of the $^{129}$Xe nucleus.
The $^{87}$Rb atoms in the vapor cell serve as a sensitive magnetometer to in situ read out the $^{129}$Xe magnetization.
The real-time output signal of the $^{87}$Rb magnetometer is connected to a feedback coil and generates a corresponding feedback field $B_{\textrm{fb}}$, with a rheostat in series with the coils to adjust feedback strength [Fig.\,\ref{figure1}(a), more details are presented in Supplementary Section I].
Because the $^{87}$Rb magnetometer measures both the $x$ and $y$ component of $^{129}$Xe polarization (with response $C_x$ and $C_y$ respectively), the feedback field can be expressed as $B_{\textrm{fb}}=\chi_1 P_x-\chi_2 P_y$.
Here, $\chi_1$ and $\chi_2$ represent the feedback gain associated with ``measuring $P_x$ and providing feedback in $y$'' and ``measuring $P_y$ and providing feedback in $y$'', respectively.
The values of $\chi_1$ and $\chi_2$ depend on factors such as the magnetometer response, the rheostat, and the coil coefficient.
The self-induced feedback field carries the information about the $^{129}$Xe spins and then produces a torque on the spins.
Equivalently, each single spin experiences the torque from the collective spins and its time evolution depends on the entirety.
Notably, this torque does not come from the dipole-dipole interaction between the single spin and the collective spins, but is mediated by the feedback field.

We now consider the dynamics of cooperative $^{129}$Xe spins under the self-induced feedback field.
The polarization of $^{129}$Xe in the $x$, $y$, and $z$ directions is denoted as $P_x$, $P_y$, and $P_z$ respectively.
The dynamics of cooperative $^{129}$Xe spins in the feedback circuit can be described by the Bloch equation:
\begin{equation}
\begin{aligned}
    \frac{\textrm{d}P_x}{\textrm{d}t}&= \gamma(P_y B_0 -P_z B_{\textrm{fb}})-\Gamma P_x \\
    &=(\gamma B_0+\gamma\chi_2 P_z) P_y - (\Gamma+\gamma\chi_1 P_z) P_x,
    \label{eqliou}
\end{aligned}
\end{equation}
where $\gamma$ is the gyromagnetic ratio of $^{129}$Xe, $\Gamma=1/T_2$ corresponds to the spin decoherence rate, and $T_2$ represents the intrinsic coherence time.
In this equation, we adopt the small angle approximation, treating $P_z$ as a constant.
To simplify the equation, we introduce two additional parameters, namely $\xi=\gamma\chi_1 P_z$ and $\Delta_\textrm{fb}=\gamma\chi_2 P_z$.
The parameter $\xi$, associated with the process of ``measuring $P_x$ and providing feedback in $\hat{y}$'', represents the modification of decoherence induced by feedback (incoherent effect).
On the other hand, the parameter $\Delta_\textrm{fb}$, linked to the process of ``measuring in $P_y$ and providing feedback in $\hat{y}$'', describes a feedback-induced frequency shift (coherent effect).
The rheostat controls the amplitude of both $\xi$ and $\Delta_\textrm{fb}$, while the sign is determined by the connecting polarity of the feedback coil.
The ratio $\Delta_\textrm{fb}/\xi$ remains constant and is determined by the $^{87}$Rb magnetometer.

\begin{figure*}[t]  
	\makeatletter
\centering
	\def\@captype{figure}
	\makeatother
	\includegraphics[scale=1]{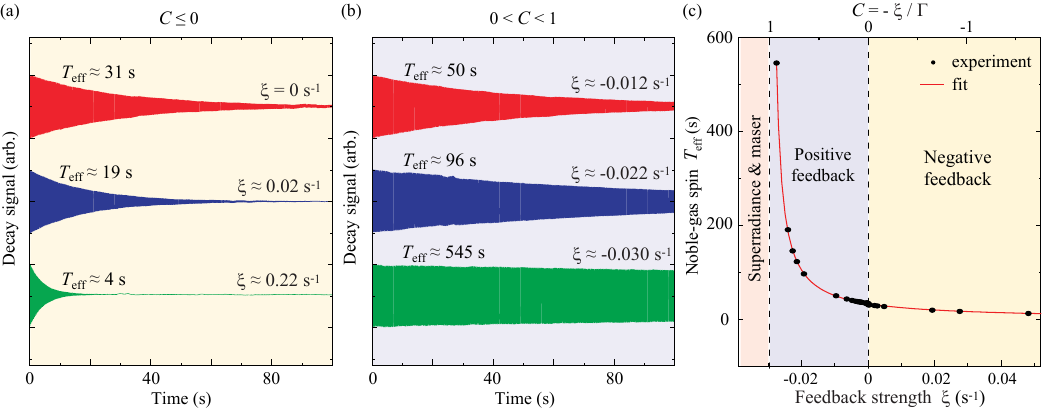}
	\caption{Demonstration of cooperative $^{129}$Xe dynamics with different feedback strengths. (a) In the regime where $C\le0$, the coherence decay rate becomes higher as $\xi$ increases. (b) In the regime where $0<C<1$, the coherence decay rate becomes smaller as $|\xi|$ increases. All the curves have been normalized and offset along the y-axis for clarity. (c) The effective coherent time $T_{\textrm{eff}}$ versus the feedback strength $\xi$. The red line shows the fit with the inverse function. $T_{\textrm{eff}}$ cannot be defined in the $C>1$ regime. Instead of exponentially decayed signal, superradiance-shaped pulses and maser occur in such $C>1$ regime.}
	\label{figure2}
\end{figure*}

We show that the cooperative spin coherence time can be significantly enhanced through manipulating the feedback strength.
According to Eq.\,(\ref{eqliou}), the decoherence rate modified by the feedback $\xi$ becomes
\begin{equation}
    \frac{1}{T_\textrm{eff}}=\Gamma+\xi
    \label{eq3},
\end{equation}
where $T_{\textrm{eff}}$ is the effective coherence time.
In order to clearly illustrate that the behaviors of the spins are closely connected with relation between $\Gamma$ and $\xi$, we define the parameter $C=-\xi/\Gamma$.
In our analysis, we focus solely on the $\chi_1$ component, disregarding the contribution of $\chi_2$ which primarily induces a frequency shift.
For $0 < C < 1$ (positive feedback), we demonstrate that spins, initially dephased from the collective spin due to random noise, exhibit a tendency to refocus towards the collective spin [Fig.\,\ref{figure1}(b)].
In the presence of the feedback field, each spin experiences a torque parallel to the collective spin, compelling them to rotate until they realign with the collective spin (Supplementary Section II).
As a result, unlike in independent dephasing scenarios, the cooperative spins are able to correct their precession phase according to the entirety, leading to an extended coherence time.
Conversely, when $C < 0$ (negative feedback), the feedback-induced torque is anti-parallel to the collective spin [Fig.\,\ref{figure1}(c)].
Under this torque, the dephased spins tend to spread out until they align in the opposite direction, effectively canceling the collective spin.
As a consequence, the decoherence rate worsens with the presence of feedback.
It is this modulation of the decoherence process that distinguishes cooperative systems from independent systems.

We demonstrate cooperative $^{129}$Xe spin dynamics by adjusting feedback parameter $\xi$.
When $\xi$ is set in the $C\le 0$ and $0<C<1$ regime, the transverse magnetization decays exponentially with modified rate.
To track changes in the coherence time,
we apply a transverse pulse to tilt the $^{129}$Xe spins at a small angle about 5$^{\circ}$ and record the resu0ltant decay signal.
The signals are fitted by exponentially-decayed sinusoidal function to determine the corresponding coherence time.
In the $C\le 0$ regime, the coherence time decreases from 31\,s to 4\,s with increasing $\xi$ [Fig.\,\ref{figure2}(a)].
In the $0<C<1$ regime, the coherence signal decays slower for larger $|\xi|$, and realizes $T_\textrm{eff}>T_2$ [Fig.\,\ref{figure2}(b)].
In our experiment, the coherence time $T_\textrm{eff}$ can be tuned to about 545\,s, which is more than one order longer than that observed without feedback ($\approx$31\,s).
Furthermore, Figure\,\ref{figure2}(c) shows the effective $^{129}$Xe coherence time for different values of $\xi$, which can be well fitted with the theoretical inverse function.
When $\xi$ is set in the $C>1$ regime, superradiance-shaped pulses and maser occur instead of exponentially decayed signal, and $T_\textrm{eff}$ can no longer be defined in such regime.

\begin{figure}[t]  
	\makeatletter
\centering
	\def\@captype{figure}
	\makeatother
	\includegraphics[scale=0.95]{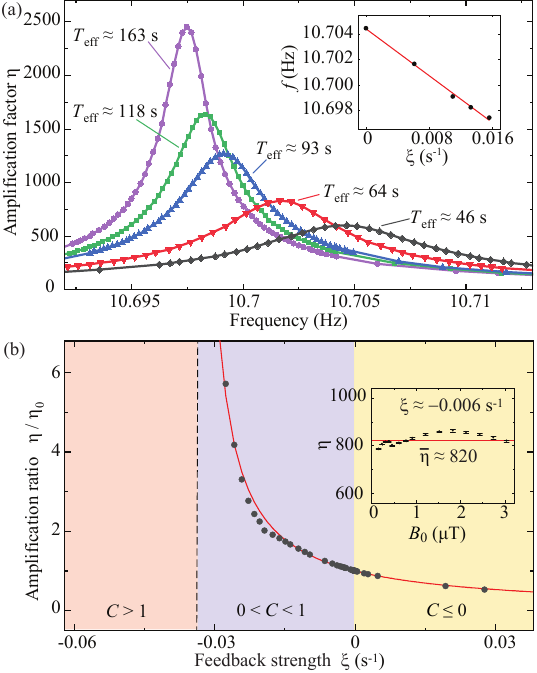}
	\caption{Demonstration of cooperative spin amplification. (a) Amplification factor swept near resonance. The data fitted well with Lorentz profile, with its maximum proportional to the effective coherent time. The inset shows a frequency shift linear to $\xi$. The fitted slope is about -0.46. (b) Plot of relative amplification factor for different feedback strength to that without feedback. The red line shows the fitting to Eq.\,(\ref{eq3}). The bias field $B_0$ is set to be about 900\,nT, corresponding to Larmor frequency 10.7\,Hz. The inset shows that amplification factor (black dots) is independent of the bias field, indicating that the amplification is available for other frequency.}
	\label{figure4}
\end{figure}


Significant magnetic-field amplification is observed using cooperative $^{129}$Xe spins.
A transverse oscillating magnetic field $\mathbf{B}_\textrm{ac}$ is applied on $^{129}$Xe spins and generates transverse magnetization of $^{129}$Xe; the magnetization induces an effective magnetic field $\mathbf{B}^\perp_\textrm{eff}$ through Fermi-contact collisions with $^{87}$Rb atoms.
As reported in Refs.\,[\onlinecite{Jiang2021}, \onlinecite{su2021search}, \onlinecite{jiang2022floquet}], the amplitude of $\mathbf{B}^\perp_\textrm{eff}$ can be significantly larger than that of $\mathbf{B}_\textrm{ac}$ with an amplification factor $\eta_0=|\mathbf{B}^\perp_\textrm{eff}|/|\mathbf{B}_\textrm{ac}|$.
The factor is determined by $\eta_0=\frac{\lambda}{2} M_0 P_0 \gamma T_2$, where $T_2$ is the intrinsic coherence time.
Such amplifiers are based on independent $^{129}$Xe spins, and their amplification ranges from 20-200 [\onlinecite{Jiang2021}, \onlinecite{su2021search}, \onlinecite{jiang2022floquet}].
In contrast, our approach enhance the coherence time with the cooperative $^{129}$Xe spins as demonstrated, leading to a modified cooperative amplification (Supplementary Section III)
\begin{equation}
    \eta=\frac{\lambda}{2} M_0 P_0 \gamma T_\textrm{eff},
    \label{eq5}
\end{equation}
where the coherence time is $T_\textrm{eff}$ instead of the intrinsic $T_2$.
This provides new opportunities to realize improved spin amplification.
We experimentally measure $\eta$ and the bandwidth of the amplifier by sweeping frequency around $^{129}$Xe resonance and recording signal response [Fig.\,\ref{figure4}(a)].
The fitting curve of Lorentz profile is overlaid on the experimental data.
We further investigate $\eta$ under different $T_\textrm{eff}$ by tuning $\xi$ and show that the resonance peak becomes narrower and higher with longer $T_\textrm{eff}$.
For example, when $T_\textrm{eff}$ is tuned to be about 163\,s, the amplification $\eta$ reaches approximately 2500.
We also find that the resonance frequency $f$ deviates from Larmor frequency $f_0$ in the presence of the feedback field [see inset of Fig.\,\ref{figure4}(a)].
As derived in Supplementary Section II, the shift $(f-f_0)$ linearly depends on $\xi$ and its slope equals to $-C_y/C_x$.
The fitted result is $f-f_0\approx-0.46\xi$.

The relative amplification $\eta / \eta_0$ is shown in Fig.\,\ref{figure4}(b).
The cooperative response leads to a 5-fold enhancement in the relative amplification $\eta / \eta_0$.
Further enhancement of $\eta$ is realized when $\xi$ approaches $-\Gamma$.
However, in practice, the fluctuation of $^{87}$Rb magnetometer response or feedback circuit resistance limits the precision of $\xi$ and makes $^{129}$Xe spins leave the amplification regime $0<C<1$.
The inset of Fig.\,\ref{figure4}(b) shows $\eta$ values under different bias field $B_0$ from 0.08\,$\mu$T to 3\,$\mu$T with $\xi\approx0.006$\,s$^{-1}$, where the amplification factor $\eta$ is nearly independent of $B_0$ and its average is about 820.
In contrast to spin-exchange-relaxation-free magnetometers that require the operation at near-zero fields below 100\,nT [\onlinecite{allred2002high}], the present $^{129}$Xe cooperative sensor can be operated in $\mu$T-level magnetic field.

\begin{figure}[t]  
	\makeatletter
\centering
	\def\@captype{figure}
	\makeatother
	\includegraphics[scale=0.95]{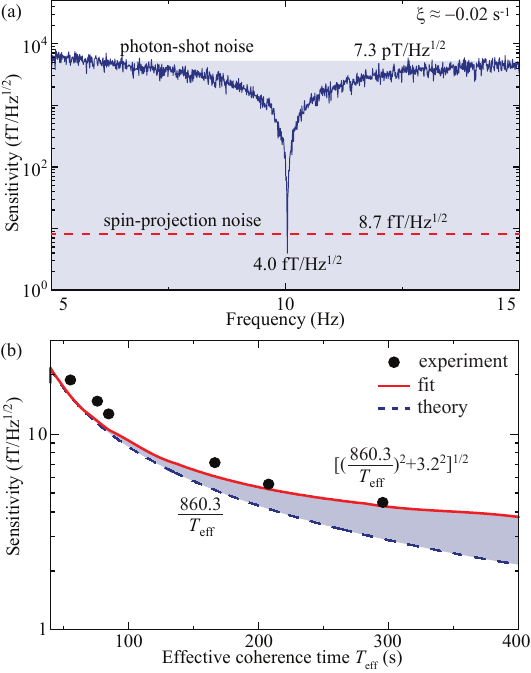}
	\caption{Sensitivity and noise analysis. (a) Sensitivity of cooperative spin-based amplifier. The cooperative response greatly improve sensitivity around the resonant point, and the best sensitivity we achieved is 4.0\,fT/Hz$^{1/2}$. The achieved sensitivity surpasses photon-shot noise limit of 7.3\,pT/Hz$^{1/2}$ and Rb spin-projection noise limit of 8.7\,fT/Hz$^{1/2}$. (b) Plot of sensitivity against $T_{\textrm{eff}}$ values. Red line shows the fit of experimentally achieved sensitivity against $T_{\textrm{eff}}$ with the model $[(a/T_\textrm{eff})^2+b^2]^{1/2}$. The sensitivity is inversely proportional to $T_{\textrm{eff}}$ when the impact of magnetic noise is mitigated, shown by the dashed line.}
	\label{figure5}
\end{figure}

As a first application, we use cooperative spin amplification to realize magnetic-field precision measurements with a fT/Hz$^{1/2}$-level sensitivity.
As an example, the bias field is set to $B_0 \approx 850$\,nT,
corresponding to $^{129}$Xe Larmor frequency $f_0 \approx 10.03$\,Hz.
By tuning the feedback strength, the effective coherence time is set to $T_\textrm{eff} \approx 300$\,s.
A resonant oscillating field $B_\textrm{ac} \approx 13.8$\,pT along the $y$ direction is applied as a test field.
Benefiting from cooperative $^{129}$Xe amplification, the applied test field is pre-amplified into 65\,nT.
By taking the response of the cooperative spin amplifier into account,
the magnetic sensitivity of $^{87}$Rb magnetometer is effectively enhanced to about 4.0\,fT/Hz$^{1/2}$ around resonance frequency, as illustrated in Fig.\,\ref{figure5}(a).
The sensitivity is over 1800 times better than photon-shot noise limit ($\approx$7.3\,pT/Hz$^{1/2}$) of $^{87}$Rb magnetometer.
Moreover, it surpasses spin-projection noise ($\approx$8.7\,fT/Hz$^{1/2}$) of $^{87}$Rb magnetometer by 2.2 fold (Supplementary Section IV).

Figure\,\ref{figure5}(b) depicts the magnetic-field sensitivity with various feedback strengths that correspond to different enhanced coherence time $T_\textrm{eff}$.
The sensitivity data are fitted with the function $[(a/T_\textrm{eff})^2+b^2]^{1/2}$, where the coefficients are estimated to be $a\approx860.3$ and $b\approx3.2$.
Here the first term originates from non-magnetic photon-shot noise, which is not amplified and can be suppressed by the amplifier.
The second term denotes real magnetic noise about 3.2\,fT/Hz$^{1/2}$ that can be amplified by the cooperative amplifier, including magnetic-shield Johnson noise and unavoidable feedback circuit magnetic noise.
As one can see, the current sensitivity is dominantly limited by the magnetic noise, which can be suppressed by existing techniques.
For example, magnetic-shield Johnson noise can be minimized by using ferrite shielding \cite{kornack2007low}.
The theoretical sensitivity is indicated by the dashed line when the potential magnetic noise is removed, e.g. the sensitivity can be improved to better than 1\,fT/Hz$^{1/2}$ when $T_\textrm{eff}$ is tuned to 900~s.
Further improvement of the cooperative amplifier can be implemented with smaller coefficient $a$, which requires high noble-gas number density, noble-gas spin polarization, and alkali-metal magnetometer response.
Extrapolating the present results to devices with alkali-noble-gas pairs with smaller spin-destruction cross section such as K-$^3$He, $a$ should be reduced to about 80 with 3\,atm $^3$He.
$^3$He spins also possess longer intrinsic coherence time ($\approx$1000\,s), which can be hours-long after enhanced by cooperative approach.
These methods would extend sensitivity below 0.1\,fT/Hz$^{1/2}$.

~\

\noindent
\textbf{Discussions.}
We would like to emphasize the main difference between this work and Fermi-contact enhancement.
First, the Fermi-contact enhancement factor $\lambda$ constitutes just a fraction of amplification factor $\eta$.
It should be noted that many other parameters are also important to realize a significant amplification factor, such as $P_0$ and $T_\textrm{eff}$.
In our experiment, the polarization of $^{129}$Xe can achieve $P_0\approx0.18$ and $T_\textrm{eff}$ is tuned to more than 500\,s, both of which is essential to realize an amplification factor of more than three orders.
Second, we introduce cooperative amplifier to further increase the amplification.
A 5-fold enhancement of $\eta$ is achieved through tuning the feedback strength, while $\lambda$ remains unchanged.

Our technique based on cooperative spins shows potential for application in other areas, such as comagnetometry - a means to measure the precession frequency of two species of nuclei, including $^{129}$Xe-$^{131}$Xe and $^{129}$Xe-$^{3}$He [\onlinecite{bulatowicz2013laboratory}, \onlinecite{gemmel2010}].
Its ability to resist noise and systematic effects associated with the magnetic field makes it useful for searches for violation of local Lorentz invariance \cite{gemmel2010} and for new spin-dependent forces \cite{tullney2013constraints, bulatowicz2013laboratory}, inertial rotation sensing \cite{kornack2005nuclear}, etc.
By allowing for long measurement times, the persistent coherence of cooperative spins allows for high accuracy in determining the precession frequency of nuclear spins, which is proportional to the measurement time to the power of -3/2 according to Cram{\'e}r-Rao lower bound \cite{gemmel2010}.
Our cooperative approach is capable of reuniting decoherence spins and resisting magnetic field gradients, making it possible to create a new class of cooperative spin comagnetometers.
According to the experiment where the coherence time is enhanced to about 20 times longer than the independent ensemble, the frequency accuracy could be improved by two orders.
It is also reported that in the $^{129}$Xe-$^{131}$Xe isotope comagnetometer, the electric quadrupole moment of $^{131}$Xe can split into triplets due to the electric field gradient induced by the glass wall \cite{feng2020}.
These triplets may narrow down benefiting from cooperation approach, thus allowing for high precision measurements of the quadrupole splitting.

Our amplification technique has potential applications in the search for hypothetical particles theorized by various models beyond the standard model, such as axions and dark photons \cite{Jiang2021, caputo2021dark}.
These particles are expected to interact with standard model particles (such as nuclear spins) and produce an oscillating pseudo-magnetic field that can be amplified using our technique.
Consequently, the search sensitivity of axions and dark photons can be significantly enhanced, leading to new empirical constraints.
With our current experimental parameters, one-day measurement yields the search sensitivity of axion dark matter $|g_\textrm{aNN}| \leq 10^{-10}$\,GeV$^{-1}$, which surpasses the most stringent supernova constraints \cite{1978JETPL, Raffelt2008} by about two orders of magnitude.
The constant $g_\textrm{aNN}$ characterizes axion-neutron coupling.
Our technique can also be applied to search for exotic spin-dependent interactions \cite{su2021search},  where axions serve as force mediators that couple the standard particles.
Using our current experiments, the search sensitivity is approximately one order of magnitude better than that in previous searches \cite{su2021search, wang2022limits}.

In conclusion, we have demonstrated a novel approach for enhancing quantum amplification through cooperative noble-gas spins, resulting in improved magnetic field sensitivity.
This approach should be generic to other noble gas, as well as alkali atoms and nitrogen-vacancy centers.
Notably, cooperative spin amplification can operate in the presence of  finite bias fields, eliminating the need for strict $\mu$-metal magnetic shielding.
This extended functionality facilitates applications such as exploring Schumann resonance of Earth \cite{fraser1975superconducting} and detecting geomagnetic field anomaly \cite{sheinker2009magnetic}.
In addition, the combination of cooperative spin amplification and Floquet engineering \cite{jiang2022floquet} may increase the bandwidth of amplification.

~\

\bibliographystyle{naturemag}
\bibliography{sample}

~\

\noindent
\textbf{Data availability}.

\noindent
Source data are provided with this paper. All other data that support the plots in this paper and other findings of this study are available from the corresponding author upon reasonable request.

~\

\noindent
\textbf{Code availability}.

\noindent
The code that supports the plots in this paper is available from the corresponding author upon reasonable request.

~\

\noindent
\textbf{Acknowledgement}.

\noindent
This work was supported by the Innovation Program for Quantum Science and Technology (Grant No. 2021ZD0303205),
National Natural Science Foundation of China (grants nos. 11661161018, 11927811, 12004371, 12150014, 12205296, 12274395), and Youth Innovation Promotion Association (Grant No. 2023474).

~\

\noindent
\textbf{Author contributions}.

\noindent
M.X. performed the experiments, analysed the data and wrote the manuscript.
M.J. and X.P. proposed the experimental concept, designed the experimental protocols, analysed the data and wrote the manuscript.
Y.W. contributed to the performance of the experiment, and edited the manuscript.
H.S. and Y.H. analysed the data and edited the manuscript.
All the authors contributed with discussions and checking the manuscript.

~\

\noindent
\textbf{Competing interests}.

\noindent
The authors declare no competing interests.

\end{document}